\documentclass{article}

\usepackage{arxiv}

\usepackage[utf8]{inputenc} 
\usepackage[T1]{fontenc}    
\usepackage{hyperref}       
\usepackage{url}            
\usepackage{booktabs}       
\usepackage{amsfonts}       
\usepackage{nicefrac}       
\usepackage{microtype}      
\usepackage{lipsum}
\usepackage{graphicx}
\graphicspath{ {./images/} }

\usepackage{amsmath}
\usepackage{lipsum}
\usepackage{verbatim}
\usepackage[T1]{fontenc}
\usepackage[utf8]{inputenc}
\usepackage[switch]{lineno}
\usepackage{upquote}
\usepackage[hang]{footmisc}
\usepackage{graphicx}
\usepackage{footnote}
\usepackage{array}
\usepackage{color}
\usepackage{framed}
\usepackage{xcolor}
\usepackage{mdframed}
\usepackage{url}
\usepackage[normalem]{ulem}
\usepackage{color}
\usepackage[caption=false]{subfig}

\title{A Passive Circuit-Emulator for a Current-controlled Memristor}

\author{
 Leonardo Barboni \\
  IIE-Fing-Udelar\\
  Montevideo, Uruguay\\
  Julio Herrera y Reissig 565 CP 11.300,  \\
  \texttt{lbarboni@fing.edu.uy} \\
}

\begin{document}
\maketitle
\begin{abstract}
A memristor is an electrical element, which has been conjectured in 1971 to complete the lumped circuit theory. Currently, researchers use memristors emulators through diodes and other passive (or active) elements to study circuits with possible attractors, chaos, and ways of implementing nonlinear transformations for low-voltage novel computing paradigms. However, to date, such passive memristor emulators have been voltage-controlled. In this study, the first circuit realization of a current-controlled passive emulator is established. The formal theory  and simulations validate the proposed circuit.
\end{abstract}


\section{Introduction}
\label{sec:introduction}

\noindent A memristor is an electrical two-terminal passive nonlinear resistance element that exhibits a well-known pinched hysteresis loop at the origin of the voltage-current plane when any bipolar-periodic zero-mean excitatory voltage or current of any value is applied across it. However, there are disagreements regarding whether a memristor can be considered a fundamental element and whether its dynamic is purely electromagnetic, as originally conjectured, or if there are other mechanisms involved such as ionic transport. Despite the controversy surrounding the technological realization, the amount of research into the properties of the pinched hysteresis loop continues to increase. Irrespective of whether the memristor is implemented by emulating its behavior through circuitry composed of other active or passive components, new studies continue to generate and sustain optimistic expectations in the scientific community about the use and advantages of the memristor. The research interest in the memristor is motivated by its promising potential for building novel integrated circuits and computing systems, as has been proposed in ~\cite{Mazumder2015,Hamdioui2017,Selmy2017}. Memristors allow the memory and time-varying processing of information through nonlinear transformations in a unique passive element.  
\\
\noindent Though some hypothetical memristor building processes appear to be compatible with those used for CMOS (layers of metals, polysilicon, and doped silicon), currently the design of building blocks with memristors is no longer acceptable owing to the restrictions of the technology design rules. Therefore, there is neither an implementation nor a proof-of-concept for a memristor-CMOS system-on-chip. Owing to the absence of integrated implementation capability and with an increasing need to better understand a pinched hysteresis attractor, researchers implement memristor emulators through diodes and other passive (or active) elements. Such emulators allow to study (theoretically and numerically) possible attractors and ways of implementing nonlinear transformations for novel computing paradigms. The first voltage-controlled memristor emulator has been proposed in \cite{Corinto2012}, and was based on a diode-bridge with a parallel $R-C$ filter as a load. Other studies~\cite{Chen2015,  QXu2016,ChenBaoEntropy2014,Njitacke2016,Xu2017,Bao2018} used voltage-controlled memristor emulators as vital building blocks of other circuits for the in-depth study of their bifurcations and chaotic behaviors. Memristors exhibit three characteristics for any bipolar periodic signal excitation, including: \textit{i)} pinched  hysteresis loop in the voltage-current plane, \textit{ii)} the area of the hysteresis loop decreases and shrinks to a single-valued V-I function when the signal excitation frequency tends to infinity, and \textit{iii)} for voltage-controlled generalized memristive time-invariant systems, the following equations apply: 

\begin{equation}
  \begin{cases}
    i_m=G(\mathbf{x},V_m)V_m  \quad \textrm{and} \quad G(\mathbf{x},0)\neq \infty \quad \forall \mathbf{x}\\ 
    \displaystyle \frac{d\mathbf{x}}{dt}=f(\mathbf{x},V_m)  \quad \textrm{where $\mathbf{x}$ represents the inner state variables}
  \end{cases}
  \label{eq:7gMCLx}
\end{equation}

\noindent Where $i_m$ is the current across the memristor, and $V_m$ is the voltage across the terminals, $G(\mathbf{x},V_m)$ is bounded, and $f(\mathbf{x},i_m)$ is the equation of state, which must be also bounded to guarantee the existence of a solution $\mathbf{x}(t)$. The area of the lobes, shapes, and orientation of the hysteresis loop evolve with frequency. All of the above mentioned references developed voltage-controlled memristor emulators (i.e. equations such as Eq.~\ref{eq:7gMCLx}). However, in this study, we implement the first-ever built passive current-controlled memristor emulator for which the following equations apply: 

\begin{equation}
  \begin{cases}
    V_m=R(\mathbf{x},i_m)i_m  \quad \textrm{and} \quad R(\mathbf{x},0)\neq \infty \quad \forall \mathbf{x}\\ 
    \displaystyle \frac{d\mathbf{x}}{dt}=f(\mathbf{x},i_m)  \quad \textrm{where $\mathbf{x}$ represents the inner state variables}
  \end{cases}
  \label{eq:jnhbvsdcsdcd}
\end{equation}

\section{Proposed current-controlled memristor}
\label{sec:proposed}

\begin{figure}[!htb]
  \begin{center}
    \includegraphics[width=0.3\textwidth, angle=0]{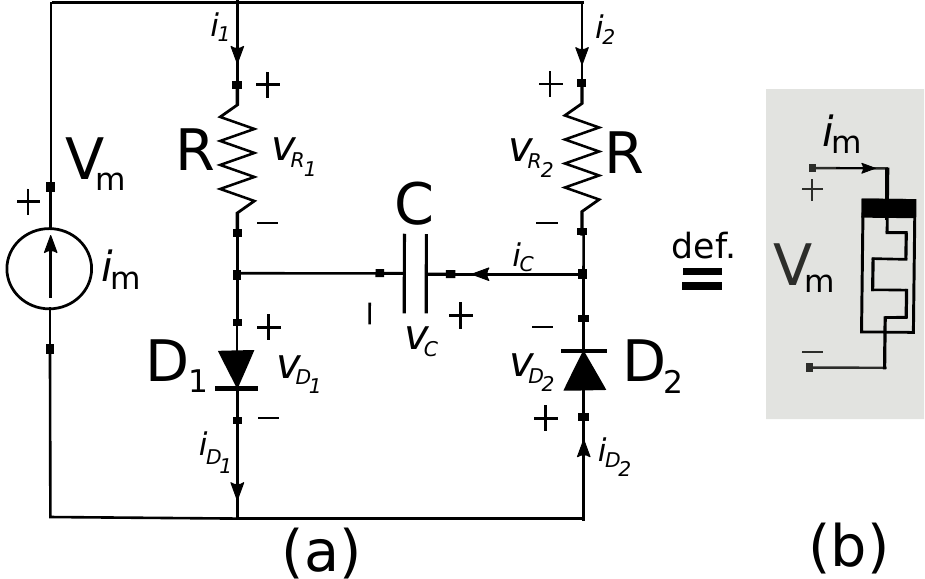} 
  \end{center}
  \caption{ \textit{(a)} Proposed current-controlled memristor circuit emulator, \textit{(b)} generalized symbol of the memristor device.}
  \label{fig:Acs3ewwregsc36}
\end{figure}

\noindent The first circuit realization of a current-controlled passive emulator is established in Fig.~\ref{fig:Acs3ewwregsc36}. It uses two diodes, two resistors, and one capacitor. 

\noindent We construct the equations by beginning with the diode equation (without considering intrinsic parasitic and high-frequency effects that produce unwanted dynamic effects), 

\begin{equation}
    i_{D_1}= I_s(e^{2\alpha V_{D_1} }-1)   \quad \textrm{and} \quad i_{D_2}= I_s(e^{ 2\alpha V_{D_2} }-1)
\label{eq:pfgzuvsZKMWkB8zJ}
\end{equation}

\noindent where $2\alpha= 1/nV_T$ and $I_s$ denote the reverse saturation current, $n$ is the emission coefficient, and $V_T$ is the thermal voltage. According to the voltage drop,

\begin{equation}
    V_{m}= Ri_{1}+V_{D_1}  \quad \textrm{and} \quad V_{m}= Ri_{2} -V_{D_2} 
\label{eq:7gMCLx4Ntf7WUJgn}
\end{equation}

\begin{equation}
    -V_{c}= V_{D_1}+V_{D_2} 
\label{eq:cDqRsqDR5qvgmc}
\end{equation}

\noindent The current $i_{m}$ corresponds to $i_{m}=i_{D_1}-i_{D_2}$. Using Eqs.~\ref{eq:pfgzuvsZKMWkB8zJ} and~\ref{eq:7gMCLx4Ntf7WUJgn},

\begin{equation}
    i_{m}= I_s \displaystyle  e^{ \alpha (V_{D_1}+V_{D_2}) } \left( \displaystyle  e^{ \alpha (V_{D_1}-V_{D_2}) }-e^{ -\alpha (V_{D_1}-V_{D_2})} \right)
\label{eq:2LK6by5B3AbBQuYs}
\end{equation}

\noindent Using Eq.~\ref{eq:cDqRsqDR5qvgmc}, it becomes convenient to express, 

\begin{equation}
    i_{m}=2 I_s e^{ - \alpha V_{c} } sinh \left( \alpha (V_{D_1}-V_{D_2})  \right)
\label{eq:dUcD}
\end{equation}

\noindent However, from Eq.~\ref{eq:7gMCLx4Ntf7WUJgn}

\begin{equation}
   2V_{m}= Ri_{m}+V_{D_1}-V_{D_2} 
\label{eq:aslplaszq}
\end{equation}

\noindent Then, the $\{V_{m}-i_{m}\}$ relation is provided,

\begin{equation}
   V_{m}= \displaystyle \frac{1}{2} Ri_{m}+ \frac{1}{2 \alpha }sinh^{-1}\left( \displaystyle \frac{i_{m}}{2I_s}  e^{\alpha V_{c} }  \right)
\label{eq:bvNv7p8szmMNKrSP}
\end{equation}

\noindent Next, we focus on the state equation. By taking $V_{m}=V_{D_1}+V_{C}+ Ri_2$ and $V_{m}=-V_{D_2}-V_{C}+ Ri_1$ we obtain,

\begin{equation}
    V_{D_1}+V_{D_2}+2V_{C}+ R(i_2-i_1) =0
\label{eq:6QcJCC6VX}
\end{equation}

\noindent Because $i_2-i_1=i_2-2i_1+i_1=i_{m}-2i_1$, and $-V_{c}= V_{D_1}+V_{D_2}$, from Eq.~\ref{eq:6QcJCC6VX} we obtain 

\begin{equation}
    i_1= \frac{V_c}{2R}+\frac{i_{m}}{2}
\label{eq:Jmk28tzZAN5z6AVc}
\end{equation}

\noindent Thus, the yield expression is, 

\begin{equation}
    i_{c}= C \frac{dV_{c}}{dt}= i_{D_1}-i_{1} \Rightarrow C \frac{dV_{c}}{dt}= i_{D_1} - \frac{V_{c}}{2R} -\frac{i_{m}}{2}
\label{eq:eSrz9TwqZL96nVA3}
\end{equation}

\noindent Now, to complete the state equation, we have to calculate $i_{D_1}$, which is straightforward (please note that according to Eq.~\ref{eq:jnhbvsdcsdcd}, $\mathbf{x}=V_c$),  

\begin{equation}
    \frac{V_{m}-V_{D_1}}{R}= i_{1} = \frac{V_{c}}{2R}+\frac{i_{m}}{2}  \Rightarrow V_{D_1}=V_{m}-\frac{Ri_{m}}{2} - \frac{V_{c}}{2}       
\label{eq:Sv2U3ZZyH67yfJdx}
\end{equation}

\noindent Therefore, from Eq.~\ref{eq:eSrz9TwqZL96nVA3} and $i_{D_1}=I_s( \displaystyle e^{2 \alpha V_{D_1} }-1)$ we obtain:

\begin{equation}
    C\frac{dV_{c}}{dt}= I_s \left( e^{ 2 \alpha V_{m} } e^{-\alpha Ri_{m}} e^{-\alpha V_c} -1 \right) -\frac{i_{m}}{2} - \frac{V_{c}}{2R}
\label{eq:2Dza8hhJCUY6ZrCC}
\end{equation}

\noindent By introducing Eq.~\ref{eq:bvNv7p8szmMNKrSP}, we obtain,

\begin{equation}
    C\frac{dV_c}{dt}= I_s e^{sinh^{-1}  \left( \displaystyle \frac{i_{m}}{2 I_s} e^{\alpha V_c} \right) -\alpha V_c }- {I_s} - \frac{i_{m}}{2} - \frac{V_c}{2R}
\label{eq:2DzaZrvsdvcx654CC}
\end{equation}

\noindent Equation ~\ref{eq:2DzaZrvsdvcx654CC} is the best evaluated Taylor series of $e^{x}$, therefore after algebraic manipulations, the state equation can be expressed as follows,

\begin{equation}
   C\frac{dV_c}{dt}=I_s \sum_{n=1}^{ \infty} \frac{1}{n!}  \displaystyle { \left( sinh^{-1}  \left( \displaystyle \frac{i_{m}}{2I_s} e^{\alpha V_c} \right) - \alpha V_c \right)}^{n}                   - \frac{i_{m}}{2} - \frac{V_c}{2R}
   \label{eq:2acvvmdfsfd254C}
\end{equation}

\noindent Finally, this memristor dynamic can be written as follows,

\begin{equation}
  \begin{cases}
    V_{m}= \displaystyle \frac{1}{2} Ri_{m}+ \frac{1}{2 \alpha }sinh^{-1}\left( \displaystyle \frac{i_{m}}{2I_s}  e^{\alpha V_{c} }  \right) \\
    \displaystyle \frac{dV_c}{dt}= \displaystyle \sum_{n=1}^{ \infty} \frac{I_s  { \left( sinh^{-1}  \left( \frac{i_{m}}{2I_s} e^{\alpha V_c} \right) - \alpha V_c \right)}^{n} }{Cn!}                   - \frac{i_{m}}{2C} - \frac{V_c}{2RC}
  \end{cases}
  \label{eq:dnascvsdwefcsaq}
\end{equation}

\noindent According to Eq.~\ref{eq:jnhbvsdcsdcd}, $V_c$ represents the inner state variable; however, it should be noted that the form $V_{m}=R(V_c,i_{m})i_{m} $  is  not achieved (i.e. it does not contain $i_{m}$ proportionality). Instead, it can be naturally achieved through division by $i_{m}$ as follows,

\begin{equation}
  \begin{cases}
    V_{m}= \displaystyle \left [ \frac{1}{2} R+ \frac{1}{2 \alpha i_{m} }sinh^{-1}\left( \displaystyle \frac{i_{m}}{2I_s}  e^{\alpha V_{c} }  \right) \right] i_{m}  \quad \textrm{for}  \quad i_m \neq 0  \\
    V_{m}=0  \quad \textrm{for} \quad  i_m = 0
    \\
    \displaystyle \frac{dV_c}{dt}= \displaystyle \sum_{n=1}^{ \infty} \frac{I_s  { \left( sinh^{-1}  \left( \frac{i_{m}}{2I_s} e^{\alpha V_c} \right) - \alpha V_c \right)}^{n} }{Cn!}                   - \frac{i_{m}}{2C} - \frac{V_c}{2RC}
  \end{cases}
  \label{eq:kmdkssdcwsd93ndb}
\end{equation}

\noindent Where according to Eq.~\ref{eq:jnhbvsdcsdcd} the term $R(V_c,i_m)$ is, 

\begin{equation}
  \begin{cases}
    R(V_c,i_m)= \displaystyle \frac{1}{2} R+ \frac{1}{2 \alpha i_{m} }sinh^{-1}\left( \displaystyle \frac{i_{m}}{2I_s}  e^{\alpha V_{c} }  \right)   \quad \textrm{for}  \quad i_m \neq 0  \\
    R(V_c,i_m)=0  \quad \textrm{for} \quad  i_m = 0
  \end{cases}
  \label{eq:sdcsdfvfs}
\end{equation}

\noindent Please, note that $R(V_c,i_m)\longrightarrow 0$ for $i_{m}\longrightarrow 0$ thus, it can be continuously extended to $i_m=0$ (the limit exists and can be obtained by L'Hospital's rule). Figure~\ref{fig:dfdbvd+3ddbsc} shows in advance an example of the zero-crossing $V-i$ (next section discusses the validation by simulation).

\section{Validation by simulation}
\label{sec:simulation}

\begin{figure}[!htb]
  \begin{center}
    \includegraphics[width=0.5\textwidth, angle=0]{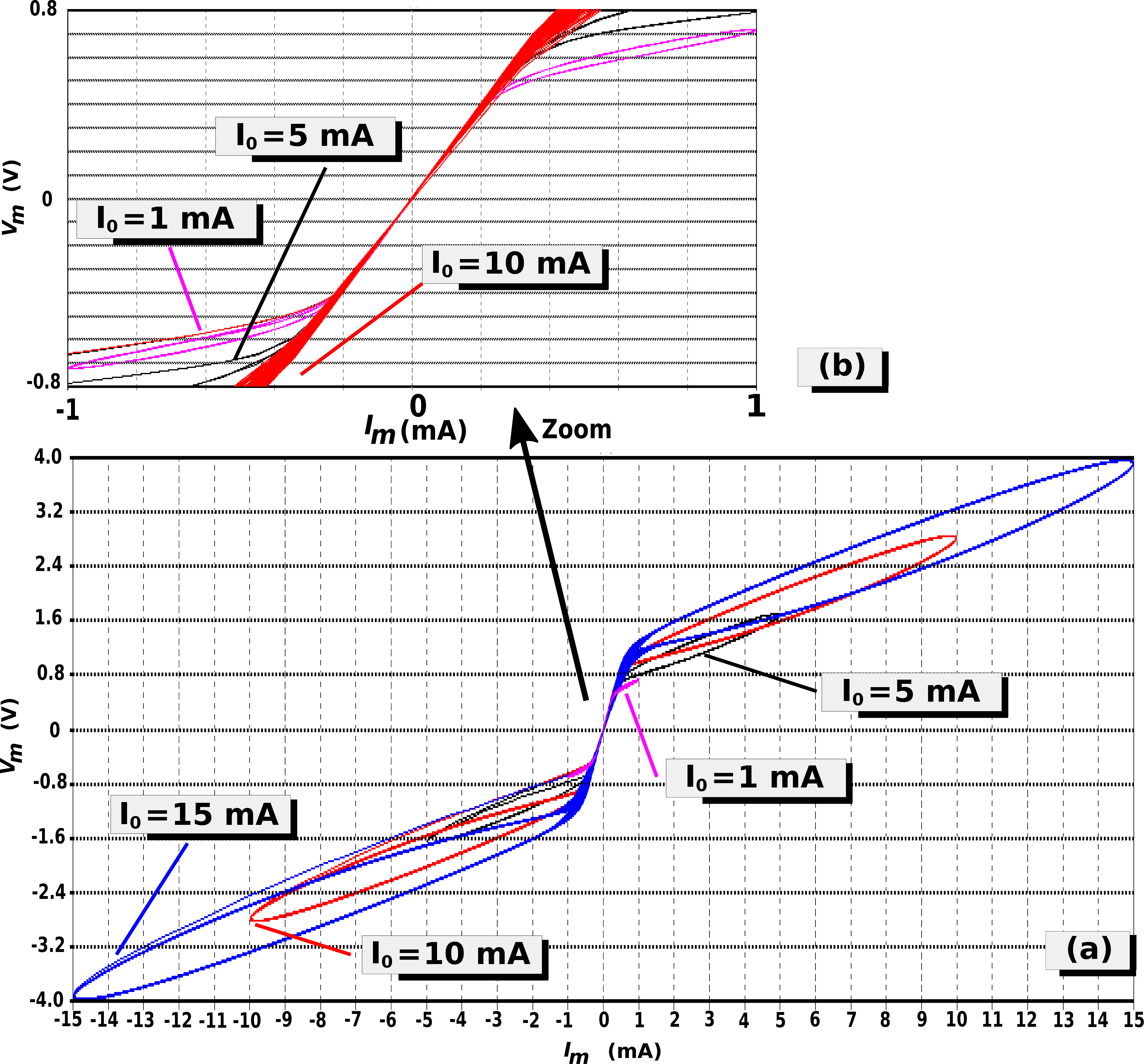} 
  \end{center}
  \caption{Current-voltage characteristics obtained through the PSpice simulations of the memristor emulator driven by a current source $i_{m}=I_osin(2\pi ft)$ at a constant frequency $f=300\;Hz$ with different $I_o$ values marked in the figure \textit{a)}: $I_o=1\;mA$, $I_o=5\;mA$, $I_o=10\;mA$ and $I_o=15\;mA$, \textit{b)} an amplified view of the zero-crossing $V-i$  omitting  the lobe at $I_o=15\;mA$ to avoid the overwhelming superposition of curves.}
  \label{fig:dfdbvd+3ddbsc}
\end{figure}

\begin{figure}[!htb]
  \begin{center}
    \includegraphics[width=0.5\textwidth, angle=0]{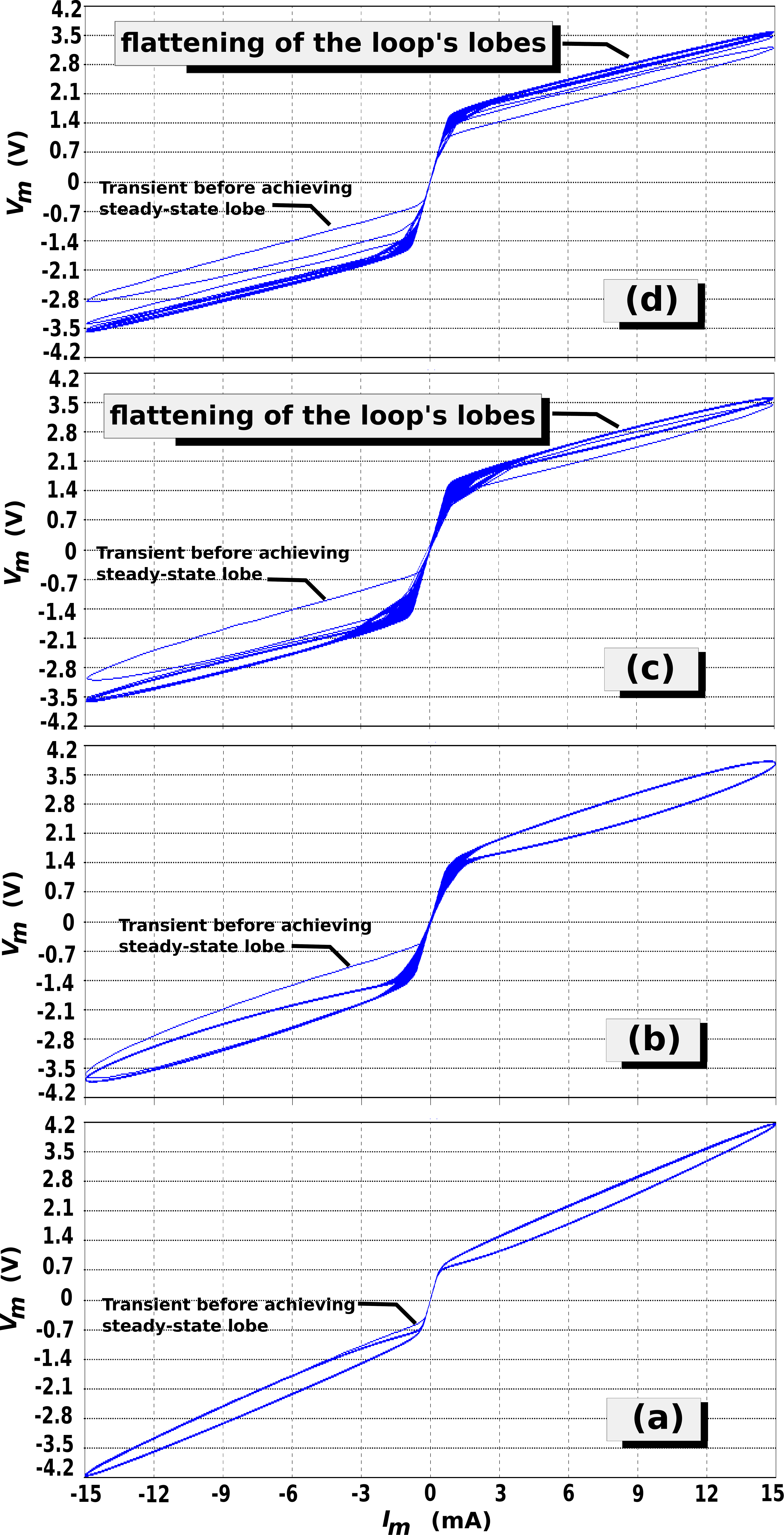} 
  \end{center}
  \caption{Simulated pinched hysteresis loop of the memristor emulator driven by a current source $i_{m}=I_osin(2\pi ft)$ with a constant amplitude $I_o=15\;mA$ at different $f$ values: \textit{a)} $f=100\;Hz$, \textit{b)} $f=500\;Hz$, \textit{c)} $f=1500\;Hz$ and \textit{d)} $f=3000\;Hz$.}
  \label{fig:123dmdnfjk3e6ogks}
\end{figure}

\noindent The following parameters were used to simulate the memristor circuit emulator proposed in Fig.~\ref{fig:Acs3ewwregsc36}: $R=270\Omega$, $C=0.5 \mu F $. The assigned diode was 1N4148 with a PSpice-model card \textit{.model 1N4148 D(Is=2.52n Rs=.568 N=1.752 Cjo=4p M=.4 tt=20n Iave=200m Vpk=75 mfg=OnSemi type=silicon)}~\cite{Standarddio}. The initial state condition $V_c=0$ was selected. The simulator used was LTspice~\cite{LTspice}. For such diode models, the magnitudes of circuit parameters and the values of input signal amplitudes and frequencies are maintained similar to compare the lobe shapes obtained in other works such as~\cite{Corinto2012,ChenBaoEntropy2014,Xu2017,Bao2018}. The current-voltage characteristics obtained from PSpice simulations are shown in Figs.~\ref{fig:dfdbvd+3ddbsc} and~\ref{fig:123dmdnfjk3e6ogks}.
 
\noindent The following observations can be made from the simulation results. The loci in the V–i plane has hysteresis loops pinched at zero in the periodic steady-state. The hysteresis loop shrinks to a single-valued function when the frequency increases and decreases, and the shape of the memristor depends on the circuit parameters and also retains the odd-symmetry property. The qualitative difference in the lobe shapes between the cases a) and b) of Fig~\ref{fig:123dmdnfjk3e6ogks} allows us to interpret that the dynamic of the memristor can exhibit a much richer behavior at low frequencies.

\section{Conclusion}
\label{sec:conclusion}

\noindent This work shows a passive circuit current-controlled memristor emulator. It overcomes the lack of current-controlled memristor commercial devices, and it can be used as part of more sophisticated circuits. Moreover, it covers a gap in the state of the art because, currently, only passive circuits voltage-controlled memristor emulators have been developed and used.
The mathematical model of the proposed memristor emulator was derived and verified by simulations.

\bibliographystyle{unsrt}  
\bibliography{referencias}  


%



\end{document}